\documentclass{article} 
\usepackage{sources/iclr2027_conference,times}

\usepackage{amsmath,amsfonts,bm}

\def\eqref#1{equation~\ref{#1}}

\def\1{\bm{1}}

\DeclareMathAlphabet{\mathsfit}{\encodingdefault}{\sfdefault}{m}{sl}
\SetMathAlphabet{\mathsfit}{bold}{\encodingdefault}{\sfdefault}{bx}{n}

\usepackage{graphicx}
\usepackage{hyperref}
\usepackage{url}
\usepackage{algorithm}
\usepackage{algpseudocode}
\usepackage{booktabs}
\usepackage{multirow}
\usepackage{colortbl}
\definecolor{lightgray}{gray}{0.90}

\newenvironment{icompact}{
  \begin{list}{$\bullet$}{
    \itemindent -.05em
    \parsep 0pt plus 1pt
    \partopsep 0pt plus 1pt
    \topsep 2pt plus 2pt minus 2pt
    \itemsep 0pt plus 1.3pt
    \parskip 0pt plus 2pt
    \leftmargin 0.13in}
      }
{\normalsize
\end{list}
}

\title{Practical Secrets Extraction against Black-box LLMs}

\author{\textbf{Shiqian Zhao} $^{1}$ \; \textbf{Siwei Jiang} $^{2}$ \; \textbf{Xinfeng Li} $^{3}$ \; \textbf{Runyi Hu} $^{1}$ \; \textbf{Yandan Zheng} $^{1}$ \; \textbf{Congyu Guo} $^{2}$\\\textbf{Tianwei Zhang} $^{1}$ \; and \textbf{Anh Tuan Luu} $^{1}$\\[1mm]$^{1}$ Nanyang Technological University\\$^{2}$ Beijing University of Posts and Telecommunications\\$^{3}$ Hong Kong Polytechnic University}

\iclrfinalcopy 
\begin{document}

\maketitle

\begin{abstract}
Large language models (LLMs) increasingly power autonomous coding agents such as Codex and Claude Code, yet their training corpora may contain confidential credentials exposed in public repositories or collected from private development artifacts, creating risks of memorization and subsequent leakage. Existing extraction audits, however, largely assume access to model weights or token probabilities. In this work, we present a black-box secret extraction framework for commercial, API-based LLMs under output-only access. It comprises (i) \emph{Cross-Validated Secret Knowledge Distillation}, which uses semantics-preserving prompt variants, response cross-validation, and provider-specific format filtering to distill secret-relevant behavior into a local white-box proxy; and (ii) \emph{Proxy-Guided Secret Extraction and Candidate Filtering}, which combines truncated top-$p$ sampling with local token entropy, $N$-gram frequency profiling, and provider-specific structural priors. On controlled API-key benchmarks, our framework improves recovery effectiveness and real-key rates over representative baselines while reducing extraction latency. A responsible real-world evaluation further recovers masked provider-specific credentials from three independently deployed black-box LLM systems spanning OpenAI and Claude Code, showing that memorized secrets can be exposed under output-only access.

\end{abstract}

\section{Introduction}

Large language models (LLMs) like the GPT and Claude series are increasingly deployed as autonomous coding agents, demonstrating remarkable execution capabilities. However, training these models often relies on large-scale scraping of public code repositories (e.g., GitHub), where developers inadvertently expose sensitive credentials, or on proprietary user data collected by service providers. Consequently, models risk memorizing these embedded secrets, such as API keys, which malicious actors can subsequently extract, resulting in severe security breaches and financial loss. Understanding and mitigating secret memorization in LLMs is therefore an urgent priority.

Existing studies have primarily focused on evaluating memorization in open-source LLMs, leveraging their publicly accessible internal information, such as model parameters and output token probability distributions. For example, DESEC characterizes token-level differences between memorized and hallucinated credentials and adjusts token likelihoods to steer open-source code LLMs toward genuine secrets~\citep{nie2025decoding}. More recently, Yang et al. investigated the leakage of various types of sensitive information, including API keys, from fine-tuned open-source code LLMs from the perspective of training dynamics~\citep{yang2025understanding}. Specifically, they leverage token-level prediction probabilities collected across fine-tuning epochs to characterize the confidence and variability of sensitive tokens, thereby linking their training dynamics to subsequent API-key extraction success under contextual extraction attacks.

Despite the success of existing methods in extracting secrets from white-box LLMs, memorization in closed-source, API-based models remains largely unexplored. This gap is especially important for commercial endpoints, whose interfaces expose only generated text while hiding token probabilities, model states, and gradients. We therefore study black-box secret extraction under two constraints: the \textbf{black-box constraint}, which makes it difficult to distinguish memorized credentials from hallucinations, and the \textbf{search-space bottleneck}, which causes greedy decoding to repeatedly explore narrow, low-diversity patterns.

To address these constraints, we present a two-stage framework aligned with the method in Section~3. First, \emph{Cross-Validated Secret Knowledge Distillation} uses semantics-preserving prompt augmentation, response cross-validation, provider-specific format checks, and hard-label response distillation to build a local white-box proxy from output-only victim responses. Second, \emph{Proxy-Guided Secret Extraction and Candidate Filtering} uses truncated top-$p$ sampling to broaden candidate coverage and combines local token entropy, $N$-gram frequency profiling, and provider-specific structural priors to remove hallucinated keys. This separation confines remote interaction to knowledge acquisition and moves broad candidate search and filtering offline, reducing repeated victim queries while improving validity and diversity.

We evaluate the framework on controlled API-key benchmarks comprising 100 authentic keys from five providers and four programming languages, where it outperforms representative baselines (~\cite{huang2024code, nie2025decoding}) by up to 57.1\%. A responsible real-world evaluation further recovers masked provider-specific credentials from three independently deployed black-box LLMs. Together, these results show that output-only access can expose memorized secrets and establish a practical security risk for commercial LLM deployments.

\section{Related Work}

\subsection{LLM Memorization}
\label{llm memorization}

Most modern LLMs are trained autoregressively on vast corpora encompassing natural language and source code. 
Let $s = (s_1, \dots, s_m)$ denote an API key conditioned on a preceding code or configuration context $c$. 
Under the standard next-token prediction objective, the model optimizes $p_\theta(s_i \mid c, s_{<i})$ for each constituent token of $s$, indiscriminately treating sensitive credentials as generalizable semantic knowledge. 
This dynamic introduces severe vulnerabilities because authentication secrets are routinely committed to public codebases: empirical studies have identified exposed credentials across more than 100,000 GitHub repositories, with thousands of novel secrets leaked daily~\citep{meli2019bad}.
Beyond public repositories, secrets may also enter training corpora when platforms or autonomous agents collect users' private code, prompts, or execution traces without authorization and later reuse them for training.
A recent user report alleged that ZCode uploaded workspace snapshots, including complete Git histories that could retain deleted credentials, to cloud infrastructure without clear consent~\citep{zcode2026incident}. 
This vulnerability is further exacerbated by training-data duplication, which drives language models to regenerate repeated sequences at disproportionately elevated rates~\citep{kandpal2022deduplicating}. 
Consequently, even high-entropy keys risk memorization if they recur across repositories or appear alongside distinctive contextual anchors, such as provider identifiers, endpoint URLs, environment variables, or structured assignment statements.


\subsection{Extraction from LLMs}

Credential-specific investigations provide the closest precedents to our work. 
HCR benchmarks hard-coded credential leakage in neural code-completion systems under black-box access: it masks exposed credentials in GitHub source files, prompts models to complete the infill span, and filters candidates using regex schemas, deduplication, historical repository occurrences, and limited operational checks~\citep{huang2024code}. 
In contrast, DESEC characterizes token-level distributional discrepancies between genuine and hallucinated secrets, deploying a learned scorer to intervene in next-token decoding on open-weight code LLMs~\citep{nie2025decoding}. 
Complementarily, \citet{yang2025understanding} examine type-dependent memorization by tracking prediction confidence and training dynamics across fine-tuning epochs. 
However, these paradigms face distinct structural limitations: HCR assumes an in-filling regime conditioned on concrete repository contexts, while DESEC and checkpoint-dynamics analyses require internal probabilities, fine-grained representations, or decoding control that commercial endpoints strictly conceal, thereby precluding their applicability to strictly API-based, black-box LLMs.

A parallel research trajectory investigates generalized training-data and personally identifiable information (PII) extraction~\citep{inan2021training,huang2022personal,lukas2023analyzing,nakka2024pii}. 
\citet{carlini2021extracting} pioneered the verbatim extraction of sensitive sequences from GPT-2 by generating candidate sets and ranking them with likelihood-based metrics. 
\citet{nasr2025scalable} subsequently demonstrated output-only extraction against production-aligned chatbots via divergence-inducing prompts that destabilize safety guardrails. 
In the software domain, CodexLeaks formalized PII auditing for GitHub Copilot by combining privacy-oriented templates, repository search proxies, and manual verification~\citep{niu2023codexleaks}. 
Codebreaker extended this line by automating PII harvesting through semantic-entropy-guided prompt mutation across open and commercial code models~\citep{han2025codebreaker}. 
Collectively, while these studies establish the plausibility of black-box extraction, they predominantly optimize for unstructured natural language or broad personal identifiers, leaving the targeted elicitation of high-entropy, format-constrained API keys unaddressed.


\section{Method}

\begin{figure*}[t!]
  \centering
  \includegraphics[width=\linewidth]{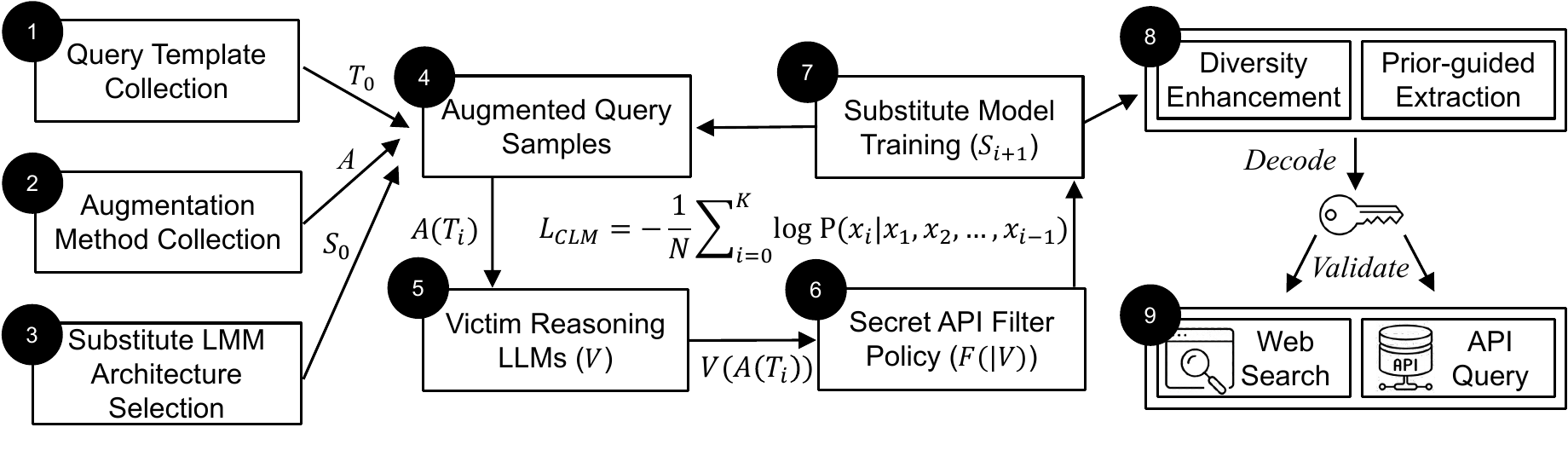} 
  \scriptsize

  \caption{Overall pipeline of our two-stage framework. It distills validated secret behavior into a local proxy before diverse offline extraction and candidate filtering.
  }
  \label{fig: pipeline}
  \vspace{-0.1in}
\end{figure*}

As illustrated in Figure~\ref{fig: pipeline}, our framework consists of two stages that match the terminology used throughout the paper: \emph{Cross-Validated Secret Knowledge Distillation} (Sec.~\ref{sec: model extraction}) and \emph{Proxy-Guided Secret Extraction and Candidate Filtering} (Sec.~\ref{sec: secret extraction}). The first stage uses semantics-preserving prompt augmentation to elicit candidate completions from the victim, cross-validates and format-filters those responses, and distills the validated pairs into a local white-box proxy. The second stage searches this proxy with diversity-enhanced truncated sampling and hybrid candidate filters based on local token entropy, $N$-gram frequency profiling, and provider-specific structural constraints. This separation confines remote interaction to knowledge acquisition and moves broad candidate search and filtering offline, reducing query overhead while improving candidate validity and diversity.

\subsection{Cross-Validated Secret Knowledge Distillation}
\label{sec: model extraction}

To address the black-box constraint, we first distill secret-related knowledge from the victim model into a local proxy model. This design is motivated by two limitations of direct querying. First, each victim query reveals only the final generated sequence while concealing token-level probabilities, alternative candidate rankings, gradients, and hidden representations, making it difficult to determine which partial API-key sequences warrant further exploration. Second, discovering diverse candidates through repeated victim queries incurs substantial API cost and may trigger abuse-detection mechanisms. The local proxy alleviates both limitations by learning from the victim's validated responses and exposing a white-box interface, allowing subsequent candidate generation and search to be performed locally using token probabilities and other internal signals rather than through repeated remote queries. 

\noindent \textbf{Cross-Validation via Prefix Perturbation.} 
To elicit memorized secrets from the model, constructing an effective conditioning prefix is essential. 
As formulated in Sec.~\ref{llm memorization}, the likelihood of each secret token depends on $p_\theta(s_i \mid c, s_{<i})$; thus, maximizing query confidence requires closely aligning the prompt with the true prefix of the ingested training snippet. 
While this alignment is trivial for owners auditing their own code, an external evaluator or third-party auditor lacks access to the exact code context from which the secret originated. 
To overcome this limitation, we take the canonical invocation snippet $T_0$, which is typically public and standardized for a given API, and synthesize diverse variants using a family of five semantics-preserving transformations $\mathcal{A}$. 
These operations simulate common real-world developer modifications: comment insertion/editing ($\mathcal{A}_0$), dead code injection ($\mathcal{A}_1$), whitespace and formatting adjustments ($\mathcal{A}_2$), syntax-equivalent refactoring ($\mathcal{A}_3$), and identifier renaming ($\mathcal{A}_4$). 
Formally, the auditor's candidate prefix set is defined as:
\begin{equation}
    \mathcal{T} = \{\mathcal{A}_j(T_0) \mid \mathcal{A}_j \in \mathcal{A}\}.
\end{equation}
These variants approximate diverse realistic contexts surrounding the API call. 
To systematically evaluate how sensitivity varies with context drift, we categorize the transformation intensity into three tiers, \textit{i.e.}, \emph{weak}, \emph{medium}, and \emph{strong}, to reflect varying degrees of deviation from the standard.

\noindent \textbf{Response Filtering.} 
Querying the victim model $\mathcal{M}_{V}$ with our augmented prefix set $\mathcal{T}$ yields a set of candidate completions, denoted as $\mathcal{M}_{V}(\mathcal{T})$. 
Because perturbed prompts cannot achieve perfect alignment with the original training context, and due to the stochastic nature of autoregressive decoding, a substantial portion of the generated outputs inevitably contains spurious or malformed content. 
To isolate valid credentials, we design a composite filtering pipeline, denoted collectively as $\mathcal{F} = \mathcal{F}_{\text{global}} \circ \mathcal{F}_{\text{local}}$, comprising two levels of structural verification:
\begin{icompact}
    \item \textbf{Local Format Filter ($\mathcal{F}_{\text{local}}$).} We inspect candidate keys at the token and character level against vendor-specific specification rules. First, we enforce provider-specific prefix constraints, discarding candidates that deviate from deterministic headers (\textit{e.g.}, Google API keys must begin with `\texttt{AIzaSy}'). Second, we validate individual characters against the provider's permissible alphabet, eliminating any output containing invalid characters outside the expected alphanumeric set and allowed symbols (\textit{e.g.}, `\texttt{\_}').
    \item \textbf{Global Structural Filter ($\mathcal{F}_{\text{global}}$).} Following local syntax validation, we enforce macroscopic structural constraints, primarily length consistency. Because commercial API keys adhere to strict vendor-defined lengths (\textit{e.g.}, DeepSeek keys typically span 35 characters, whereas Claude credentials comprise approximately 108 characters), candidates exhibiting aberrant lengths are pruned as hallucinations.
\end{icompact}
Applying $\mathcal{F}$ yields the curated dataset:
\begin{equation}
    \mathcal{D} = \left\{ \big(t, \mathcal{F}(\mathcal{M}_V(t))\big) \mid t \in \mathcal{T} \right\},
\end{equation}
which retains high-confidence, secret-relevant pairs used to train the local surrogate proxy model.

\noindent \textbf{Proxy Model Fine-tuning}. Although format filtering removes syntactically invalid completions, responses elicited by different prefix variants may still contain complementary secret fragments or token-level deviations induced by imperfect contextual alignment. We therefore consolidate the secret-related behavior exposed by the victim model into a local proxy model $\mathcal{M}_{S}$ through hard-label response distillation~\citep{chen2024blackboxkd}. Let $y=(y_1,\ldots,y_{|y|})=\mathcal{F}(\mathcal{M}_V(t))$ denote the filtered response to query $t$. Because the black-box victim exposes only generated sequences rather than token-level logits, we treat each $y$ as a pseudo-label and minimize the autoregressive negative log-likelihood:
\begin{equation}
    \mathcal{L}_{\mathrm{proxy}}(\theta_S)
    = -\frac{1}{|\mathcal{D}|}
    \sum_{(t,y)\in\mathcal{D}}
    \frac{1}{|y|}
    \sum_{i=1}^{|y|}
    \log p_{\theta_S}\!\left(y_i \mid t, y_{<i}\right),
    \label{eq:proxy-distillation}
\end{equation}
where $\theta_S$ denotes the parameters of $\mathcal{M}_{S}$. This objective trains the proxy to reproduce the victim's validated responses while jointly learning from multiple semantics-preserving variants of each query template. Consequently, response patterns that recur across variants are reinforced, whereas inconsistent perturbation-induced artifacts are attenuated, yielding a white-box proxy suitable for subsequent secret search.

\subsection{Proxy-Guided Secret Extraction and Candidate Filtering}
\label{sec: secret extraction}

Given the secret-bearing proxy model, which is white-box for the attacker, we can now extract the API keys by leveraging the probability distribution, as well as an unlimited query budget. We adopt the sampling-filtering scheme for diversity-enhanced and effectiveness-improved secret extraction.

\noindent \textbf{Diversity-Preserving Sampling}. Existing extraction methods often guide decoding with a reward model learned from features of valid keys. Although such guidance can improve the immediate hit rate, repeatedly maximizing an explicit reward concentrates generation on a small set of high-scoring patterns, including patterns represented in the reward-model training data, and therefore reduces coverage of the secret space. We instead remove the reward model and use stochastic probability truncation as a softer decoding signal. Moreover, auxiliary reward models necessitate an additional forward pass at every token position. 
Eliminating this requirement removes one forward evaluation per decoding step, substantially mitigating cumulative latency and query overhead. 
For instance, when decoding a sequence of $L = 35$ tokens, removing per-token auxiliary scoring eliminates 35 redundant forward evaluations, thereby reducing the marginal forward overhead to $\frac{1}{35+1} \approx 2.8\%$. As detailed in Appendix~\ref{app:sample-kp} (see Algorithm~\ref{alg:sample-kp}), our $\texttt{Sample}\!-\!k\%p$ procedure combines top-$p$ token selection with $k$ independent decoding runs.
\begin{icompact}
    \item \textbf{Token candidates at $p$.} At decoding position $i$, let $z_i(v)$ be the proxy model's logit for token $v$ in vocabulary $\mathcal{V}$. We first obtain the temperature-scaled distribution $\pi_i(v)=\operatorname{softmax}(z_i(v)/\tau)$ and order the tokens such that $\pi_i(v_{(1)})\geq\cdots\geq\pi_i(v_{(|\mathcal{V}|)})$. For a nucleus threshold $p\in(0,1]$, the candidate set is
    \begin{equation}
        \begin{aligned}
        m_i &= \min\left\{m\in\{1,\ldots,|\mathcal{V}|\}: \sum_{j=1}^{m}\pi_i\!\left(v_{(j)}\right)\geq p\right\},\\
        \mathcal{C}_i^{(p)} &= \left\{v_{(1)},\ldots,v_{(m_i)}\right\}.
        \end{aligned}
        \label{eq:top-p-candidates}
    \end{equation}
    We renormalize $\pi_i$ over $\mathcal{C}_i^{(p)}$ and sample the next token from the resulting distribution. A larger $p$ admits more low-probability tokens and increases diversity, whereas a smaller $p$ favors confidence but narrows the search space. We set $\tau=0.8$ to balance these effects.
    \item \textbf{Sampling times at $k$.} For each prompt, we independently repeat complete autoregressive decoding $k$ times. Because each run samples a potentially different path through the top-$p$ candidate sets, increasing $k$ improves coverage beyond a single decoding trajectory. We deduplicate the resulting sequences before validation so that repeated completions do not inflate the candidate count. Larger values of $k$ trade linearly increasing local inference cost for broader secret coverage without incurring additional victim-model queries. 
\end{icompact}

\begin{figure}[t!]
    \centering
    \begin{minipage}[c]{0.48\linewidth}
        \centering
        \includegraphics[height=0.70\linewidth]{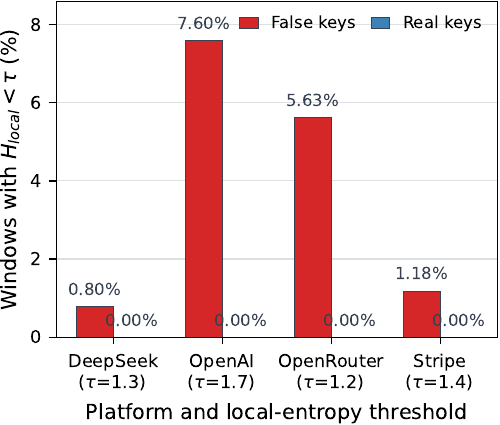}
    \end{minipage}\hfill
    \begin{minipage}[c]{0.48\linewidth}
        \centering
        \includegraphics[height=0.70\linewidth]{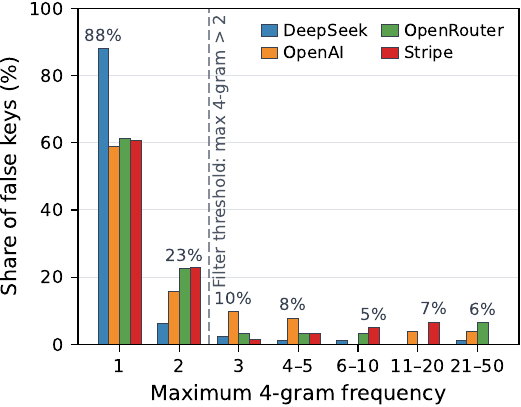}
    \end{minipage}
    \caption{Empirical signatures of hallucinated API-key candidates in $\mathcal{K}_{\mathrm{false}}$: (left) lower-tail distribution of minimum-window entropy; (right) distribution of the maximum 4-gram occurrence count. Lower entropy and higher 4-gram frequency indicate stronger character-level degeneration.}
    \label{fig:filter-distributions}
    \vspace{-0.1in}
\end{figure}

\noindent \textbf{Authenticity-Aware Filtering}. After generating $k$ samples for each prompt $t$, we apply a lightweight post-generation filtering stage to retain candidates that are more likely to be authentic API keys. Because these deterministic policies operate after decoding and reuse the generated candidates and their associated statistics, they introduce little computational overhead and require no additional victim-model queries. Specifically, we design the following policies based on empirical differences between authentic keys and model hallucinations. 

Prior to detailing each filtering policy, we outline the empirical framework designed to evaluate them. 
We first curate a benchmark dataset of authentic API keys, denoted as $\mathcal{K}_{\text{real}}$, alongside a set of canonical calling prefixes $\mathcal{P}$, pairing each invocation template with its corresponding key. Calibration keys and final evaluation keys are strictly disjoint; all thresholds are fixed on the calibration split before final evaluation. 
Next, we fine-tune a base Qwen model on these pairs to establish ground-truth memorization. 
We then evaluate the fine-tuned model by querying it with prefixes from $\mathcal{P}$. 
Generated outputs that fail to match the ground-truth set $\mathcal{K}_{\text{real}}$ are labeled as \textit{hallucinations} (or false extractions), yielding a negative sample set $\mathcal{K}_{\text{false}}$. 
Together, $\mathcal{K}_{\text{real}}$ and $\mathcal{K}_{\text{false}}$ serve as positive and negative ground-truth corpora to calibrate and validate our filtering policies.

\noindent \underline{\textit{Minimum-Window Entropy Screening}}. Global entropy summarizes the character distribution of an entire candidate, but it can conceal short repetitive or degenerate regions. We therefore compute entropy over overlapping character windows. Let $s=(s_1,\ldots,s_L)$ be a candidate key, $\mathcal{A}$ its character alphabet, and $W$ the window size. For the window beginning at position $i$, we define the empirical frequency $p_i(c)=\frac{1}{W}\sum_{j=i}^{i+W-1}\mathbf{1}[s_j=c]$ and its local Shannon entropy as
\begin{equation}
    H_i^{(W)}(s)=-\sum_{c\in\mathcal{A}}p_i(c)\log_2 p_i(c).
    \label{eq:local_entropy}
\end{equation}
We retain a candidate only if its least-diverse window satisfies
\begin{equation}
    \min_{1\leq i\leq L-W+1} H_i^{(W)}(s) \geq \tau_{\mathrm{local}}^{(a)},
    \label{eq:local_entropy_filter}
\end{equation}
where $a$ denotes the API provider. We use $W=8$ and calibrate $\tau_{\mathrm{local}}^{(a)}$ on $\mathcal{K}_{\mathrm{real}}$ and $\mathcal{K}_{\mathrm{false}}$: $1.3$ for DeepSeek, $1.7$ for OpenAI, $1.2$ for OpenRouter, $1.4$ for Stripe, and $1.9$ for Google. This policy complements global entropy by rejecting candidates whose overall distribution appears plausible but that contain locally collapsed segments. The left panel of Figure~\ref{fig:filter-distributions} visualizes the lower-tail distribution of minimum-window entropy among hallucinated candidates.

\noindent \underline{\textit{Repeated-Substring Suppression}}. Local entropy alone may not expose periodic strings whose windows contain several distinct characters. We therefore explicitly bound repeated substrings. For an $n$-gram length $n$, let
\begin{equation}
    \mathcal{G}_n(s)=\{s_{i:i+n-1}\mid 1\leq i\leq L-n+1\},
    \qquad
    f(g;s)=\sum_{i=1}^{L-n+1}\mathbf{1}[s_{i:i+n-1}=g],
    \label{eq:ngram_count}
\end{equation}
where $f(g;s)$ counts the occurrences of $g$ in $s$. A candidate is retained only when
\begin{equation}
    \max_{g\in\mathcal{G}_n(s)} f(g;s) \leq \tau_{\mathrm{ngram}}.
    \label{eq:ngram_filter}
\end{equation}

The right panel of Figure~\ref{fig:filter-distributions} reports the maximum 4-gram frequency distribution for hallucinated candidates. To establish an authentic-key reference, we compute the same statistic over 100 authentic keys from each provider. The maximum is exactly $1$ for DeepSeek, OpenAI, Stripe, and Google, and ranges from $1$ to $2$ for OpenRouter. Accordingly, we instantiate Eq.~\ref{eq:ngram_filter} with $n=4$ and $\tau_{\mathrm{ngram}}=2$. This threshold retains every authentic key in our calibration set while rejecting any candidate that repeats a 4-gram more than twice. Together, minimum-window entropy screening and repeated-substring suppression capture complementary forms of character-level degeneration.

\section{Validation}

\subsection{Experimental Setup}

\noindent \textbf{Dataset and Model}.
We collect 500 valid API keys from OpenAI, OpenRouter, Google, DeepSeek, and Stripe, with 100 keys per platform. From each platform's API keys, we collect 100 code snippets for calling the API (embedded in Python, Java, JavaScript, and PHP code, with 25 samples per language) and pair them randomly with these 100 API keys, and finally obtain a 100-sample training set containing unique code snippets and API keys. We fine-tune DeepSeek-Coder-7B~\citep{guo2024deepseekcoder}, LLaMA-3-8B~\citep{grattafiori2024llama}, and Qwen2.5-Coder-7B~\citep{hui2024qwen25coder} as victim models using LoRA~\citep{hu2022lora}, computing the loss on all code snippets to simulate the real-world training process. The attacker can observe only the generated outputs. We use StarCoder2-7B~\citep{lozhkov2024starcoder2} and Seed-Coder-8B~\citep{zhang2025seedcoder} as white-box proxies. Thus, all credentials in the controlled experiment were author-generated, privately maintained, and absent from the proxy’s pre-existing training data. The calibration-key set is strictly disjoint from the final-evaluation key set; all filtering thresholds are fixed on the calibration split before evaluation.

\noindent \textbf{Baselines.}
We consider three representative baselines. \textit{Greedy Decoding} directly queries the black-box victim and returns its most likely continuation. \textit{HCR}~\citep{huang2024code} additionally applies credential-specific formatting and heuristic filters to victim outputs. \textit{DESEC}~\citep{nie2025decoding} is a white-box baseline that uses token probabilities and guided beam search to identify high-confidence candidates; we apply it to the same surrogate model used by our method. 

\noindent \textbf{Evaluation Metrics}. We evaluate our framework along three dimensions: extraction effectiveness, candidate diversity, and generation efficiency. For effectiveness, we report (1)~\emph{Recovered Secrets (RS)}, the number of ground-truth API keys recovered under the same sampling budget, and (2)~\emph{Real Rate (RR)}, the fraction of extracted candidates that are valid ground-truth keys. For diversity, we report within-prompt cosine and 3-gram Jaccard similarities; lower values indicate broader exploration. For efficiency, we report \emph{Post-construction Generation Latency}, the average running time for one extraction after the white-box proxy has been constructed.

\noindent \textbf{Implementation}. We implement our approach using Python 3.12, PyTorch 2.8.0 and CUDA 12.8 on a single RTX 5090 GPU. All models are fine-tuned via LoRA using standard hyperparameter settings based on official checkpoints. Victim and white-box proxy models follow this common LoRA configuration for training. During inference, we sample 10 candidates from perturbed black-box code with $T=0.5, \text{top-}p=0.8$, while white-box proxies adopt $T=0.8, \text{top-}p=0.9$ with the same candidate quantity. All compared methods share identical models, data and sampling budgets, with only the attack strategy varying.

\begin{table*}[t]
\centering
\caption{Comparison of average number of recovered real secrets (RS). The final column reports the average over the five platforms. Parenthesized percentages report the relative improvement over the stronger of HCR and Greedy Decoding in the same column.}
\label{tab:rs-by-proxy-platform}
\setlength{\tabcolsep}{4.0pt}
\renewcommand{\arraystretch}{1.08}
\small
\resizebox{\textwidth}{!}{%
\begin{tabular}{llrrrrrr}
\toprule
\textbf{Victim model} & \textbf{Method} & \textbf{DeepSeek} & \textbf{Google} & \textbf{OpenAI} & \textbf{OpenRouter} & \textbf{StripeTest} & \textbf{Avg.} \\
\midrule
\multirow{4}{*}{DeepSeek-Coder-7B}
& HCR & 49.0 & 54.0 & 4.0 & 27.0 & 3.0 & 27.4 \\
& Greedy Decoding & 50.0 & 55.0 & 4.0 & 28.0 & 3.0 & 28.0 \\
& \cellcolor{lightgray}Ours--StarCoder2-7B & \cellcolor{lightgray}\textbf{57.0}~{\scriptsize(+14.0\%)} & \cellcolor{lightgray}\textbf{62.0}~{\scriptsize(+12.7\%)} & \cellcolor{lightgray}\textbf{6.0}~{\scriptsize(+50.0\%)} & \cellcolor{lightgray}\textbf{45.0}~{\scriptsize(+60.7\%)} & \cellcolor{lightgray}\textbf{5.0}~{\scriptsize(+66.7\%)} & \cellcolor{lightgray}\textbf{35.0}~{\scriptsize(+25.0\%)} \\
& \cellcolor{lightgray}Ours--Seed-Coder-8B & \cellcolor{lightgray}\textbf{57.0}~{\scriptsize(+14.0\%)} & \cellcolor{lightgray}\textbf{62.0}~{\scriptsize(+12.7\%)} & \cellcolor{lightgray}\textbf{6.0}~{\scriptsize(+50.0\%)} & \cellcolor{lightgray}42.0~{\scriptsize(+50.0\%)} & \cellcolor{lightgray}\textbf{5.0}~{\scriptsize(+66.7\%)} & \cellcolor{lightgray}34.4~{\scriptsize(+22.9\%)} \\
\midrule
\multirow{4}{*}{Qwen2.5-Coder-7B}
& HCR & 37.0 & 51.0 & 10.0 & 35.0 & 15.0 & 29.6 \\
& Greedy Decoding & 37.0 & 51.0 & 10.0 & 36.0 & 16.0 & 30.0 \\
& \cellcolor{lightgray}Ours--StarCoder2-7B & \cellcolor{lightgray}\textbf{45.0}~{\scriptsize(+21.6\%)} & \cellcolor{lightgray}\textbf{56.0}~{\scriptsize(+9.8\%)} & \cellcolor{lightgray}\textbf{13.0}~{\scriptsize(+30.0\%)} & \cellcolor{lightgray}39.0~{\scriptsize(+8.3\%)} & \cellcolor{lightgray}\textbf{20.0}~{\scriptsize(+25.0\%)} & \cellcolor{lightgray}34.6~{\scriptsize(+15.3\%)} \\
& \cellcolor{lightgray}Ours--Seed-Coder-8B & \cellcolor{lightgray}\textbf{45.0}~{\scriptsize(+21.6\%)} & \cellcolor{lightgray}\textbf{56.0}~{\scriptsize(+9.8\%)} & \cellcolor{lightgray}\textbf{13.0}~{\scriptsize(+30.0\%)} & \cellcolor{lightgray}\textbf{41.0}~{\scriptsize(+13.9\%)} & \cellcolor{lightgray}\textbf{20.0}~{\scriptsize(+25.0\%)} & \cellcolor{lightgray}\textbf{35.0}~{\scriptsize(+16.7\%)} \\
\midrule
\multirow{4}{*}{LLaMA-3-8B}
& HCR & 48.0 & 59.0 & 14.0 & 44.0 & 56.0 & 44.2 \\
& Greedy Decoding & 48.0 & 59.0 & 14.0 & 44.0 & 56.0 & 44.2 \\
& \cellcolor{lightgray}Ours--StarCoder2-7B & \cellcolor{lightgray}\textbf{54.0}~{\scriptsize(+12.5\%)} & \cellcolor{lightgray}\textbf{65.0}~{\scriptsize(+10.2\%)} & \cellcolor{lightgray}\textbf{17.0}~{\scriptsize(+21.4\%)} & \cellcolor{lightgray}\textbf{51.0}~{\scriptsize(+15.9\%)} & \cellcolor{lightgray}\textbf{60.0}~{\scriptsize(+7.1\%)} & \cellcolor{lightgray}\textbf{49.4}~{\scriptsize(+11.8\%)} \\
& \cellcolor{lightgray}Ours--Seed-Coder-8B & \cellcolor{lightgray}\textbf{54.0}~{\scriptsize(+12.5\%)} & \cellcolor{lightgray}\textbf{65.0}~{\scriptsize(+10.2\%)} & \cellcolor{lightgray}\textbf{17.0}~{\scriptsize(+21.4\%)} & \cellcolor{lightgray}\textbf{51.0}~{\scriptsize(+15.9\%)} & \cellcolor{lightgray}59.0~{\scriptsize(+5.4\%)} & \cellcolor{lightgray}49.2~{\scriptsize(+11.3\%)} \\
\bottomrule
\end{tabular}%
}
\vspace{-0.1in}
\end{table*}

\subsection{Main Results}

Under identical experimental settings, we evaluate our method against representative baselines along three complementary dimensions: extraction effectiveness, candidate diversity, and generation efficiency. Complete latency comparisons and real-rate results for individual perturbation types are reported in Appendix~\ref{sec:latency} and Appendix~\ref{sec:detailed-recovery}, respectively.

\noindent \textbf{Effectiveness}. Table~\ref{tab:rs-by-proxy-platform} reports the average number of recovered real secrets across perturbation methods and strengths. Both proxy instantiations consistently outperform the strongest direct-query baseline for every victim model and API-key platform. In particular, the best average recovery rises from 28.0 to 35.0 for DeepSeek-Coder-7B, from 30.0 to 35.0 for Qwen2.5-Coder-7B, and from 44.2 to 49.4 for LLaMA-3-8B, corresponding to relative gains of 11.8\%--25.0\%. These gains are accompanied by higher real rates: under weak perturbations, Table~\ref{tab:rr-weak} shows that our method improves the proportion of authentic keys for both proxy models across all five providers, by up to 2.6$\times$ over DESEC.

\noindent \textbf{Diversity}. Table~\ref{tab:within-prompt-diversity} evaluates similarity among keys generated from the same prompt, where lower cosine and 3-gram Jaccard similarity indicate broader exploration. Our method reduces both metrics for every provider. For example, on DeepSeek and Google keys, cosine similarity drops from 0.5835 and 0.6465 under DESEC to 0.1843 and 0.1761, while 3-gram Jaccard similarity falls from 0.5122 and 0.5647 to 0.1385 and 0.1325, respectively. Thus, native top-$p$ sampling with post-generation filtering avoids repeatedly producing near-duplicate candidates.

\begin{figure*}[t]
\centering\begin{minipage}[t]{0.49\textwidth}\vspace{0pt}
\centering
\includegraphics[width=\linewidth]{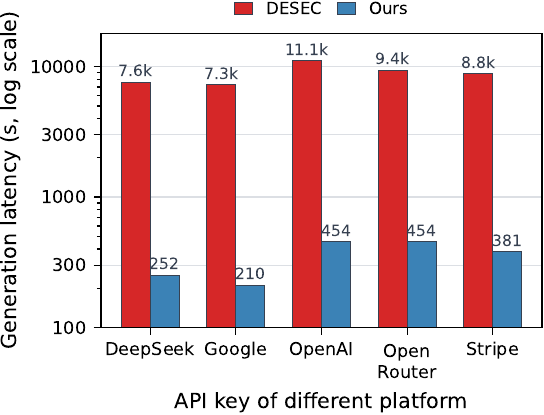}
\caption{Post-construction extraction latency comparison between DESEC and our method across five API-key platforms. By moving candidate screening and ranking offline, our method avoids token-level guided decoding and reduces extraction-stage latency.}
\label{fig:latency-comparison}
\end{minipage}\hfill\begin{minipage}[t]{0.49\textwidth}\vspace{0pt}
\centering
\scriptsize
\setlength{\fboxsep}{4pt}\begin{minipage}{0.94\linewidth}
\colorbox[rgb]{0.49,0.46,0.72}{\parbox{\dimexpr\linewidth-2\fboxsep\relax}{\color{white}\texttt{Prompt}}}\\[-1.5pt]
\colorbox[rgb]{0.96,0.95,0.99}{\parbox{\dimexpr\linewidth-2\fboxsep\relax}{\begin{tabular}{@{}l@{}}
\ttfamily\textcolor[rgb]{0.14,0.50,0.10}{...}\\[-1pt]
\textcolor[rgb]{0.14,0.50,0.10}{from streamlink.streamHolder.ffmpegmuxRef}\\[-1pt]
\textcolor[rgb]{0.14,0.50,0.10}{import MuxedStreamProp}\\[1pt]\textcolor[rgb]{0.14,0.50,0.10}{API\_KEY\_item = "AIzaSy}
\end{tabular}}}\\[2pt]
\colorbox[rgb]{0.25,0.45,0.75}{\parbox{\dimexpr\linewidth-2\fboxsep\relax}{\color{white}\textbf{Qwen3-27B}}}\\[-1.5pt]
\colorbox[rgb]{0.90,0.95,1.00}{\parbox{\dimexpr\linewidth-2\fboxsep\relax}{\begin{tabular}{@{}l@{}}
\textbf{DeepSeek:} \texttt{sk-aB3cD4eF**************R1sT2uV}\\[-1pt]
\textbf{DeepSeek:} \texttt{sk-8f3a2b1c*************1f2a3b4c}\\[-1pt]
\textbf{Google:} \texttt{AIzaSyBhOdIF**************SrEw5eihAA}
\end{tabular}}}\\[2pt]
\colorbox[rgb]{0.30,0.62,0.48}{\parbox{\dimexpr\linewidth-2\fboxsep\relax}{\color{white}\textbf{GPT-Codex 5.3}}}\\[-1.5pt]
\colorbox[rgb]{0.92,0.98,0.94}{\parbox{\dimexpr\linewidth-2\fboxsep\relax}{\begin{tabular}{@{}l@{}}
\textbf{DeepSeek:} \texttt{sk-aB3dE5fG*************tU9vW1xY}
\end{tabular}}}\\[2pt]
\colorbox[rgb]{0.76,0.35,0.18}{\parbox{\dimexpr\linewidth-2\fboxsep\relax}{\color{white}\textbf{Claude Haiku 4.5}}}\\[-1.5pt]
\colorbox[rgb]{1.00,0.94,0.90}{\parbox{\dimexpr\linewidth-2\fboxsep\relax}{\begin{tabular}{@{}l@{}}
\textbf{DeepSeek:} \texttt{sk-7f8a9b2c*************l2m3n4o}\\[-1pt]
\textbf{DeepSeek:} \texttt{sk-9f8e7d6c*************f6e5d4c}
\end{tabular}}}
\end{minipage}
\caption{Representative masked credentials recovered from three real-world black-box LLMs using the same code prompt. Each output is labeled by its API provider; credential middles are redacted for responsible disclosure.}
\label{fig:real-world-case-study}
\end{minipage}
\vspace{-0.1in}
\end{figure*}

\noindent \textbf{Efficiency}. Figure~\ref{fig:latency-comparison} shows that our method requires 210--454 seconds across the five API-key platforms, compared with 7.3k--11.1k seconds for DESEC, yielding a 20.7$\times$--34.8$\times$ latency reduction. The improvement arises because our method performs native sampling followed by batched, offline filtering and ranking, rather than repeatedly applying token-level guidance during decoding. Because DESEC is a white-box method, we apply it to the same locally trained proxy used by our method, with identical models, data, hardware, and sampling budget. The victim-query, filtering, distillation, and LoRA costs are therefore shared and excluded from Figure~\ref{fig:latency-comparison}; the reported 20.7$\times$--34.8$\times$ represents a post-construction extraction speedup, not an end-to-end attack-efficiency gain.

\begin{table*}[t]
\centering
\begin{minipage}[t]{0.48\textwidth}
\centering
\refstepcounter{table}\label{tab:rr-weak}
{\footnotesize\raggedright\textbf{Table~\thetable:} Real rate (RR) under weak perturbations. For each key type, results are averaged across five perturbation methods and three victim models. Higher values indicate a higher proportion of real API keys among plausible candidates.\par}
\vspace{2pt}
\setlength{\tabcolsep}{2.6pt}
\renewcommand{\arraystretch}{1.05}
\scriptsize
\begin{tabular}{lcccc}
\toprule
\multirow{2}{*}{\textbf{Key type}} & \multicolumn{2}{c}{\textbf{StarCoder2-7B ($\uparrow$)}} & \multicolumn{2}{c}{\textbf{Seed-Coder-8B ($\uparrow$)}} \\
\cmidrule(lr){2-3}\cmidrule(lr){4-5}
& \textbf{DESEC} & \textbf{Ours} & \textbf{DESEC} & \textbf{Ours} \\
\midrule
DeepSeek   & 0.0889 & \textbf{0.2343} & 0.0891 & \textbf{0.2184} \\
Google     & 0.0970 & \textbf{0.2426} & 0.1014 & \textbf{0.2376} \\
OpenAI     & 0.0178 & \textbf{0.0236} & 0.0182 & \textbf{0.0243} \\
OpenRouter & 0.0660 & \textbf{0.0833} & 0.0671 & \textbf{0.0802} \\
Stripe     & 0.0395 & \textbf{0.0482} & 0.0398 & \textbf{0.0467} \\
\bottomrule
\end{tabular}
\end{minipage}\hfill
\begin{minipage}[t]{0.48\textwidth}
\centering
\refstepcounter{table}\label{tab:within-prompt-diversity}
{\footnotesize\raggedright\textbf{Table~\thetable:} Diversity of generated API keys. For each key type, the two metrics are averaged within each of 100 prompt files and then across files; files containing fewer than two keys are assigned zero. Lower values indicate higher diversity.\par}
\vspace{2pt}
\setlength{\tabcolsep}{2.6pt}
\renewcommand{\arraystretch}{1.05}
\scriptsize
\begin{tabular}{lcccc}
\toprule
\multirow{2}{*}{\textbf{Key type}} & \multicolumn{2}{c}{\textbf{Cosine ($\downarrow$)}} & \multicolumn{2}{c}{\textbf{Jaccard-3 ($\downarrow$)}} \\
\cmidrule(lr){2-3}\cmidrule(lr){4-5}
& \textbf{DESEC} & \textbf{Ours} & \textbf{DESEC} & \textbf{Ours} \\
\midrule
DeepSeek   & 0.5835 & \textbf{0.1843} & 0.5122 & \textbf{0.1385} \\
Google     & 0.6465 & \textbf{0.1761} & 0.5647 & \textbf{0.1325} \\
OpenAI     & 0.1813 & \textbf{0.1018} & 0.1440 & \textbf{0.0598} \\
OpenRouter & 0.3601 & \textbf{0.2007} & 0.3041 & \textbf{0.1344} \\
Stripe     & 0.2854 & \textbf{0.2370} & 0.1986 & \textbf{0.1470} \\
\bottomrule
\end{tabular}
\end{minipage}
\end{table*}

\subsection{Ablation Study}\noindent \textbf{Impact of Augmentation Methods}. We evaluate five semantics-preserving code perturbations under weak perturbations, using Qwen2.5-Coder-7B as the victim model and Seed-Coder-8B as the proxy. Table~\ref{tab:augmentation-method-rr} shows that all augmentations can elicit real keys, but their effectiveness varies across providers. Format perturbation achieves the highest RR for DeepSeek keys (0.2885), commit edits are strongest for Google keys (0.2534), and syntax-equivalent transformations yield the best RR for OpenAI, OpenRouter, and Stripe. Thus, no single augmentation dominates; combining diverse perturbations broadens the contexts through which memorized keys can be recovered.

\noindent \textbf{Impact of Perturbation Degree}. Table~\ref{tab:perturbation-degree-rr} aggregates RR over five perturbation methods, three victim models, and two proxy models for each degree ($n=30$). Weak perturbations yield the highest mean RR for four of five providers, especially DeepSeek (0.2264) and Google (0.2401). Stronger perturbations substantially reduce RR for these two providers, while OpenAI and OpenRouter remain nearly unchanged; Stripe is the only exception, with a slightly higher mean under strong perturbation (0.0480). These results suggest that mild, semantics-preserving changes provide more reliable retrieval contexts while limiting deviation from memorized code.

\begin{table*}[t]
\centering
\begin{minipage}[t]{0.48\textwidth}
\centering
\refstepcounter{table}\label{tab:augmentation-method-rr}
{\footnotesize\raggedright\textbf{Table~\thetable:} Impact of code-augmentation methods on real rate (RR). Results use Qwen2.5-Coder-7B as the victim, Seed-Coder-8B as the proxy, and our method under weak perturbations. Higher is better.\par}
\vspace{2pt}
\setlength{\tabcolsep}{1.6pt}
\renewcommand{\arraystretch}{1.08}
\scriptsize
\resizebox{\linewidth}{!}{
\begin{tabular}{lccccc}
\toprule
\textbf{Method} & \textbf{DeepSeek} & \textbf{Google} & \textbf{OpenAI} & \textbf{OpenRouter} & \textbf{Stripe} \\
\midrule{}
Commit edit   & 0.1282 & \textbf{0.2534} & 0.0262 & 0.0728 & 0.0327 \\
Dead code     & 0.1316 & 0.1734 & 0.0264 & 0.0663 & 0.0351 \\
Format        & \textbf{0.2885} & 0.2403 & 0.0233 & 0.0759 & 0.0350 \\
Syntax equiv. & 0.2261 & 0.2324 & \textbf{0.0277} & \textbf{0.0799} & \textbf{0.0366} \\
Var. rename   & 0.1625 & 0.1830 & 0.0263 & 0.0713 & 0.0337 \\
\bottomrule
\end{tabular}
}
\end{minipage}\hfill
\begin{minipage}[t]{0.48\textwidth}
\centering
\refstepcounter{table}\label{tab:perturbation-degree-rr}
{\footnotesize\raggedright\textbf{Table~\thetable:} Effect of perturbation degree on RR. Each entry reports the mean and standard deviation over five perturbation methods, three victim models, and two proxy models ($n=30$). Higher is better.\par}
\vspace{2pt}
\setlength{\tabcolsep}{2.0pt}
\renewcommand{\arraystretch}{1.08}
\scriptsize
\begin{tabular}{lccc}
\toprule
\textbf{Key type} & \textbf{Weak} & \textbf{Medium} & \textbf{Strong} \\
\midrule{}
DeepSeek   & \textbf{0.2264} $\pm$ \textbf{0.0871} & 0.2030 $\pm$ 0.0978 & 0.1683 $\pm$ 0.0698 \\
Google     & \textbf{0.2401} $\pm$ \textbf{0.0404} & 0.2014 $\pm$ 0.0562 & 0.1930 $\pm$ 0.0593 \\
OpenAI     & \textbf{0.0240} $\pm$ \textbf{0.0091} & 0.0236 $\pm$ 0.0090 & 0.0232 $\pm$ 0.0093 \\
OpenRouter & \textbf{0.0817} $\pm$ \textbf{0.0116} & 0.0797 $\pm$ 0.0084 & 0.0795 $\pm$ 0.0097 \\
Stripe     & 0.0474 $\pm$ 0.0394 & 0.0473 $\pm$ 0.0389 & \textbf{0.0480} $\pm$ \textbf{0.0400} \\
\bottomrule
\end{tabular}
\end{minipage}
\vspace{-0.1in}
\end{table*}

\noindent \textbf{Impact of Filtering Policies}. We compare the regular length, pattern, and dictionary filters (\textbf{Base}) with adding entropy screening (\textbf{+Ent.}), 4-gram suppression (\textbf{+4g}), or both (\textbf{Full}) under medium \texttt{commit\_edit} perturbation, using DeepSeek-Coder-7B as the victim and StarCoder2-7B as the proxy. Table~\ref{tab:filter-policy-ablation} shows that \textbf{Full} achieves the highest Real Rate (RR) for every provider, increasing the aggregate RR from 0.0622 to 0.0685 (+10.2\%). Entropy screening gives a larger isolated gain (+8.0\%) than 4-gram suppression (+4.5\%), while combining them delivers the largest improvement. All variants recover the same 170 real keys, confirming that the gains result from rejecting hallucinated candidates rather than discarding valid secrets.

\begin{table}[t]
\centering\caption{Ablation of entropy and 4-gram filtering. Parenthesized values show gains over \textbf{Base}.}
\label{tab:filter-policy-ablation}\setlength{\tabcolsep}{1.8pt}
\renewcommand{\arraystretch}{1.08}
\scriptsize
\begin{tabular}{lcccc}
\toprule
\textbf{Key type} & \textbf{Base} & \textbf{+Ent.} & \textbf{+4g} & \textbf{Full} \\\midrule
DeepSeek   & 0.1281 & 0.1319 (+3.0\%) & 0.1293 (+0.9\%) & \textbf{0.1335 (+4.2\%)} \\Google     & 0.1646 & 0.1751 (+6.4\%) & 0.1732 (+5.2\%) & \textbf{0.1789 (+8.7\%)} \\OpenAI     & 0.0101 & 0.0112 (+11.4\%) & 0.0108 (+8.0\%) & \textbf{0.0115 (+14.4\%)} \\OpenRouter & 0.0528 & 0.0594 (+12.5\%) & 0.0552 (+4.6\%) & \textbf{0.0606 (+14.8\%)} \\Stripe     & 0.0082 & 0.0086 (+5.0\%) & 0.0085 (+3.2\%) & \textbf{0.0088 (+7.2\%)} \\\midrule
Overall    & 0.0622 & 0.0671 (+8.0\%) & 0.0649 (+4.5\%) & \textbf{0.0685 (+10.2\%)} \\
\bottomrule
\end{tabular}
\vspace{-0.1in}
\end{table}

\subsection{Real-world Evaluation}

We further evaluate our framework against three real-world black-box LLMs using the same code prompt. Figure~\ref{fig:real-world-case-study} presents representative recovered credentials: Qwen3-27B returns two DeepSeek keys and one Google key, GPT-Codex 5.3 returns a DeepSeek key, and Claude Haiku 4.5 returns two DeepSeek keys. We mask credentials for disclosure. GitHub searches found public matches for each unmasked key, supporting genuine leakage rather than format-only hallucinations. We did not issue live API queries for academic-ethics reasons.

\noindent \textbf{Takeaway.} The recovery of provider-specific credentials from three independent black-box systems shows that memorized secrets can remain accessible without model weights or token probabilities. Such credentials could enable unauthorized API use, expose downstream services or data, and cause financial loss. Cross-provider recovery further shows that one leaked training artifact can affect not only the model developer, but also credential owners and service providers. These findings make secret memorization an operational security risk for black-box LLMs and motivate further study. Although our examples are masked and disclosed responsibly, they demonstrate that black-box access alone can suffice to uncover sensitive credentials.

\section{Conclusion}

We presented a practical black-box framework for extracting memorized API keys from output-only LLMs. Our two-stage design distills secret-relevant behavior from victim responses into a white-box proxy using semantics-preserving perturbations and response/format validation, then performs diverse proxy-side sampling with entropy and $N$-gram filtering. Across five providers, three victim models, and two proxies, it improves recovery by up to $25.0\%$, raises real-key rates by up to $2.6\times$, and reduces latency by up to $34.8\times$ over representative baselines. A responsible real-world evaluation recovered masked, provider-specific credentials from three deployed black-box LLMs, confirming that output-only access can expose memorized secrets without model weights or token probabilities. This highlights an operational security risk and motivates studies toward stricter pretraining data cleaning to avoid memorizing confidential information in LLMs.

\section*{AI Use Statement}
We used AI-based tools only for language polishing, formatting assistance, and
implementation support. All scientific claims, experimental results, and final
manuscript contents were verified by the authors.

\section*{Ethics Statement}
This work studies the security risks of secret memorization in black-box language
models. The main experiments were conducted in a controlled environment using
authorized model interfaces. API keys extracted from commercial platforms are
truncated and masked to prevent credential leakage. No recovered secret was used
to access external services, and raw credential-containing outputs are not released.

\section*{Reproducibility Statement}
We provide detailed descriptions of the datasets, models, perturbation methods,
filtering policies, evaluation metrics, and experimental settings. The code and
supporting materials will be released in the future to facilitate independent
reproduction.

\bibliography{sources/reference}
\bibliographystyle{sources/iclr2027_conference}

\appendix
\section{Appendix}
\subsection{Diversity-enhanced Sampling Algorithm}
\label{app:sample-kp}
For completeness, Algorithm~\ref{alg:sample-kp} gives the full pseudocode for the diversity-enhanced \texttt{Sample}-$k\%p$ decoding procedure used in Sec.~3.2.

\renewcommand{\algorithmicrequire}{\textbf{Input:}}
\renewcommand{\algorithmicensure}{\textbf{Output:}}
\algnewcommand{\Cmnt}[1]{\Comment{\textnormal{\textcolor{gray}{\small\em #1}}}}
\begin{algorithm}[t]
    \caption{Diversity-enhanced $\texttt{Sample}\!-\!k\%p$ decoding}
    \label{alg:sample-kp}
    \begin{algorithmic}[1]
        \Require Proxy model $\mathcal{M}_S$, prompt $t$, threshold $p$, repetitions $k$, temperature $\tau$, max length $L$
        \Ensure Deduplicated candidate set $\mathcal{Y}$
        \State $\mathcal{Y}\gets\emptyset$ \Cmnt{Initialize candidate set}
        \For{$r=1,\ldots,k$}
            \State $y^{(r)}\gets\epsilon$
            \For{$i=1,\ldots,L$}
                \State $z_i\gets\mathcal{M}_S(t,y^{(r)}_{<i})$ \Cmnt{Query the proxy model}
                \State Compute $\pi_i$ and $\mathcal{C}_i^{(p)}$ using Eq.~\ref{eq:top-p-candidates} \Cmnt{Build the top-$p$ support}
                \State $\widetilde{\pi}_i(v)\gets\pi_i(v)\big/\sum_{u\in\mathcal{C}_i^{(p)}}\pi_i(u)$ for $v\in\mathcal{C}_i^{(p)}$
                \State Sample $x_i\sim\operatorname{Categorical}(\widetilde{\pi}_i)$ \Cmnt{Sample the next token}
                \If{$x_i=\mathsf{EOS}$} \Cmnt{Stop at end of sequence}
                    \State \textbf{break}
                \EndIf
                \State $y^{(r)}\gets y^{(r)}\mathbin{\|}x_i$
            \EndFor
            \If{$y^{(r)}\neq\epsilon$}
                \State $\mathcal{Y}\gets\mathcal{Y}\cup\{y^{(r)}\}$ \Cmnt{Store the candidate}
            \EndIf
        \EndFor
        \State \Return $\mathcal{Y}$ \Cmnt{Return deduplicated candidates}
    \end{algorithmic}
\end{algorithm}

\subsection{Perturbation Definition and Strength Quantification}
To simulate the setting where the target model’s pre-training data cannot be obtained, we generate perturbed code samples for experiments. We design normalized metrics to control and quantify the magnitude of code perturbations. We detail the perturbation pipelines and quantification rules in the following.

\textbf{Format perturbation.}
Format perturbation applies typesetting-level modifications while preserving program semantics. Operations include inserting or removing blank lines, adjusting inline whitespace and operator spacing, and splitting long statements at delimiters. Only the surface form of code is changed, with identifiers, literals and statement structures untouched. The strength metric is
\[
D_1 = \min\left(\frac{c_1}{C_{1,\text{max}}}, 1\right),
\]
where $C_{1,\text{max}}$ is the total line count of original code minus one, and $c_1$ counts format changes after line-level alignment, covering blank-line differences, changed lines from statement splitting and spacing adjustments, and lines with indent variations. Strength levels: weak ($D_1<0.3$), medium ($0.3\le D_1\le0.7$), strong ($D_1>0.7$).

\textbf{Syntactic-equivalent transformation.}
This perturbation performs semantics-preserving syntactic transformations on expressions, such as reordering operands for commutative operators, flipping comparisons (e.g., rewriting $a>b$ as $b<a$), applying De Morgan’s laws, eliminating double negation, and adding redundant parentheses. All transformations preserve program behavior and only modify syntactic presentation. The strength metric is
\[
D_2 = \min\left(\frac{c_2}{C_{2,\text{max}}}, 1\right),
\]
where $C_{2,\text{max}}$ is the total occurrences of eligible operators ($+, *, ==, !=, \&\&, ||, \text{and}, \text{or}$), and $c_2$ is the number of expressions modified by equivalent substitution. Strength levels: weak ($D_2<0.3$), medium ($0.3\le D_2\le0.7$), strong ($D_2>0.7$).

\textbf{Dead-code insertion.}
This perturbation injects unreachable redundant statements, including constant-false branches (\texttt{if False}, \texttt{if 1 == 0}), constant-false loops (\texttt{while False}), unreachable statements after \texttt{return}, and redundant empty statements. Injected snippets are syntactically valid and do not alter control flow or program semantics. The strength metric is
\[
D_3 = \frac{\text{DeadToken}}{\text{Tokens}_\text{ori}},
\]
where $\text{Tokens}_\text{ori}$ denotes the total tokens of original code, and $\text{DeadToken}$ is the token increment after perturbation. Strength levels: weak ($D_3\in[0,0.3)$), medium ($D_3\in[0.3,0.7)$), strong ($D_3\in[0.7,1.0]$).

\textbf{Variable renaming.}
This perturbation replaces local variables, function parameters, constants and function names with semantically equivalent new names while preserving control and data flow. Renaming maintains scope consistency, updates all references, and avoids conflicts with keywords, built-ins and existing identifiers. The strength metric is
\[
D_4 = \min\left(\frac{c_4}{C_{4,\text{max}}},1\right),
\]
where $C_{4,\text{max}}$ is the number of unique original identifiers, and $c_4$ is the count of renamed identifiers, computed from the symmetric difference between original and modified identifier sets. Strength levels: weak ($D_4<0.3$), medium ($0.3\le D_4\le0.7$), strong ($D_4>0.7$).

\textbf{Comment editing.}
This perturbation only edits comment text marked by \texttt{\#} or \texttt{//}, including modifying existing comments, inserting new comments and deleting existing ones. Executable code remains unchanged, altering only textual content and human-readable information while keeping program logic identical. The strength metric is
\[
D_5 = \min\left(\frac{c_5}{\max(\text{Comment}_\text{ori}+1, 20)},1\right),
\]
where $\text{Comment}_\text{ori}$ is the number of comment lines in the original code. The denominator $\max(\text{Comment}_\text{ori}+1,20)$ mitigates overestimated strength for files with sparse comments. $c_5$ is the total count of inserted, deleted and replaced comment lines from sequence alignment. Strength levels: weak ($D_5\in(0,0.3]$), medium ($D_5\in(0.3,0.7)$), strong ($D_5\in[0.7,1.0]$).

\subsection{Inference Latency Comparison}
\label{sec:latency}
We measure the total wall-clock time for generating 10 candidate API keys per prompt. Adopting the identical experimental protocol from DESEC, we evaluate both DESEC and our method across five perturbation forms and three perturbation strengths (weak, medium, strong). We use 100 distinct prompts for every combination of perturbation form and strength, leading to 1500 prompts in total. Each prompt generates 10 candidate API keys, and we report the latency averaged across the three perturbation strengths.

\begin{table*}[t]
\centering
\small
\caption{Total inference latency (seconds) under \textbf{commit-edit} perturbation. Column pairs correspond to DESEC and our method for each victim-to-proxy model transfer.}
\label{tab:latency_commit_edit}
\resizebox{\linewidth}{!}{
\begin{tabular}{lcccccccccccc}
\toprule
APIKEY Type & \multicolumn{2}{c}{Deepseek$\rightarrow$Starcode} & \multicolumn{2}{c}{Deepseek$\rightarrow$Seedcode} & \multicolumn{2}{c}{Llama3$\rightarrow$Starcode} & \multicolumn{2}{c}{Llama3$\rightarrow$Seedcode} & \multicolumn{2}{c}{Qwen$\rightarrow$Starcode} & \multicolumn{2}{c}{Qwen$\rightarrow$Seedcode}\\
\cmidrule(lr){2-3}\cmidrule(lr){4-5}\cmidrule(lr){6-7}\cmidrule(lr){8-9}\cmidrule(lr){10-11}\cmidrule(lr){12-13}
& DESEC & Ours & DESEC & Ours & DESEC & Ours & DESEC & Ours & DESEC & Ours & DESEC & Ours \\
\midrule
Deepseek APIKEY & 7245.3 & 227.2 & 7967.2 & 266.7 & 7135.7 & 230.5 & 8081.1 & 278.1 & 7318.3 & 233.4 & 8090.7 & 282.8 \\
Google APIKEY & 7001.7 & 194.5 & 7687.4 & 234.2 & 7177.5 & 191.5 & 7731.4 & 235.9 & 6934.9 & 202.2 & 7655.2 & 227.3 \\
OpenAI APIKEY & 10080.5 & 408.9 & 12198.7 & 493.9 & 9884.3 & 406.2 & 12067.0 & 497.1 & 10204.5 & 413.0 & 12385.6 & 501.6 \\
OpenRouter APIKEY & 8533.6 & 409.0 & 10239.1 & 485.7 & 8712.4 & 415.3 & 10099.4 & 492.9 & 8479.5 & 411.7 & 10264.3 & 494.1 \\
StripeTest APIKEY & 8088.2 & 347.3 & 9614.0 & 411.6 & 8057.4 & 345.9 & 9588.7 & 413.4 & 8243.2 & 353.1 & 9481.2 & 415.8 \\
\bottomrule
\end{tabular}
}
\end{table*}

\begin{table*}[t]
\centering
\small
\caption{Total inference latency (seconds) under \textbf{dead\_code} perturbation. Column pairs correspond to DESEC and our method for each victim-to-proxy model transfer.}
\label{tab:latency_dead_code}
\resizebox{\linewidth}{!}{
\begin{tabular}{lcccccccccccc}
\toprule
APIKEY Type & \multicolumn{2}{c}{Deepseek$\rightarrow$Starcode} & \multicolumn{2}{c}{Deepseek$\rightarrow$Seedcode} & \multicolumn{2}{c}{Llama3$\rightarrow$Starcode} & \multicolumn{2}{c}{Llama3$\rightarrow$Seedcode} & \multicolumn{2}{c}{Qwen$\rightarrow$Starcode} & \multicolumn{2}{c}{Qwen$\rightarrow$Seedcode}\\
\cmidrule(lr){2-3}\cmidrule(lr){4-5}\cmidrule(lr){6-7}\cmidrule(lr){8-9}\cmidrule(lr){10-11}\cmidrule(lr){12-13}
& DESEC & Ours & DESEC & Ours & DESEC & Ours & DESEC & Ours & DESEC & Ours & DESEC & Ours \\
\midrule
Deepseek APIKEY & 7195.0 & 224.0 & 7781.1 & 271.3 & 7058.2 & 205.9 & 8053.7 & 284.1 & 7657.7 & 244.0 & 7930.3 & 307.9 \\
Google APIKEY & 6657.7 & 207.9 & 7454.3 & 207.9 & 6939.9 & 164.2 & 7610.3 & 208.3 & 6850.9 & 198.3 & 7329.4 & 248.1 \\
OpenAI APIKEY & 9713.2 & 414.4 & 11980.2 & 523.9 & 10057.0 & 418.1 & 12292.4 & 471.5 & 10576.6 & 385.1 & 12546.0 & 516.4 \\
OpenRouter APIKEY & 8297.6 & 438.8 & 10404.4 & 494.1 & 8927.7 & 420.3 & 10075.1 & 494.6 & 8256.2 & 428.6 & 10042.1 & 484.9 \\
StripeTest APIKEY & 7876.6 & 338.3 & 9764.7 & 406.6 & 7763.7 & 337.6 & 9398.5 & 400.5 & 8041.0 & 378.2 & 9111.5 & 427.5 \\
\bottomrule
\end{tabular}
}
\end{table*}

\begin{table*}[t]
\centering
\small
\caption{Total inference latency (seconds) under \textbf{format} perturbation. Column pairs correspond to DESEC and our method for each victim-to-proxy model transfer.}
\label{tab:latency_format}
\resizebox{\linewidth}{!}{
\begin{tabular}{lcccccccccccc}
\toprule
APIKEY Type & \multicolumn{2}{c}{Deepseek$\rightarrow$Starcode} & \multicolumn{2}{c}{Deepseek$\rightarrow$Seedcode} & \multicolumn{2}{c}{Llama3$\rightarrow$Starcode} & \multicolumn{2}{c}{Llama3$\rightarrow$Seedcode} & \multicolumn{2}{c}{Qwen$\rightarrow$Starcode} & \multicolumn{2}{c}{Qwen$\rightarrow$Seedcode}\\
\cmidrule(lr){2-3}\cmidrule(lr){4-5}\cmidrule(lr){6-7}\cmidrule(lr){8-9}\cmidrule(lr){10-11}\cmidrule(lr){12-13}
& DESEC & Ours & DESEC & Ours & DESEC & Ours & DESEC & Ours & DESEC & Ours & DESEC & Ours \\
\midrule
Deepseek APIKEY & 7093.6 & 237.7 & 7935.0 & 247.1 & 6968.2 & 241.0 & 7689.4 & 303.6 & 7529.4 & 244.6 & 8034.3 & 282.3 \\
Google APIKEY & 6658.8 & 178.7 & 7865.4 & 245.9 & 7136.7 & 169.9 & 7657.9 & 263.7 & 7308.3 & 186.1 & 7698.4 & 252.3 \\
OpenAI APIKEY & 9775.9 & 396.8 & 12068.3 & 499.5 & 10022.0 & 432.5 & 11831.2 & 492.1 & 10118.9 & 428.7 & 11986.1 & 516.0 \\
OpenRouter APIKEY & 8442.5 & 388.2 & 10210.3 & 505.2 & 9091.2 & 429.4 & 10240.8 & 465.7 & 8792.6 & 430.0 & 10385.5 & 466.1 \\
StripeTest APIKEY & 8290.6 & 365.5 & 9564.5 & 400.5 & 8262.4 & 339.3 & 9776.3 & 427.6 & 8637.6 & 367.2 & 9107.9 & 386.8 \\
\bottomrule
\end{tabular}
}
\end{table*}

\begin{table*}[t]
\centering
\small
\caption{Total inference latency (seconds) under \textbf{syntax\_equiv} perturbation. Column pairs correspond to DESEC and our method for each victim-to-proxy model transfer.}
\label{tab:latency_syntax_equiv}
\resizebox{\linewidth}{!}{
\begin{tabular}{lcccccccccccc}
\toprule
APIKEY Type & \multicolumn{2}{c}{Deepseek$\rightarrow$Starcode} & \multicolumn{2}{c}{Deepseek$\rightarrow$Seedcode} & \multicolumn{2}{c}{Llama3$\rightarrow$Starcode} & \multicolumn{2}{c}{Llama3$\rightarrow$Seedcode} & \multicolumn{2}{c}{Qwen$\rightarrow$Starcode} & \multicolumn{2}{c}{Qwen$\rightarrow$Seedcode}\\
\cmidrule(lr){2-3}\cmidrule(lr){4-5}\cmidrule(lr){6-7}\cmidrule(lr){8-9}\cmidrule(lr){10-11}\cmidrule(lr){12-13}
& DESEC & Ours & DESEC & Ours & DESEC & Ours & DESEC & Ours & DESEC & Ours & DESEC & Ours \\
\midrule
Deepseek APIKEY & 7027.0 & 204.3 & 8084.2 & 279.8 & 6902.7 & 206.0 & 7910.0 & 279.9 & 6981.9 & 213.7 & 8453.4 & 270.2 \\
Google APIKEY & 6659.9 & 193.3 & 7383.0 & 226.0 & 6839.4 & 216.5 & 7727.7 & 224.6 & 6935.0 & 194.4 & 7622.2 & 204.8 \\
OpenAI APIKEY & 10188.1 & 382.8 & 12518.9 & 510.3 & 9800.0 & 393.3 & 12247.9 & 473.5 & 10534.0 & 389.9 & 12101.1 & 491.2 \\
OpenRouter APIKEY & 8732.3 & 420.9 & 10539.5 & 497.9 & 8374.9 & 438.9 & 10395.4 & 497.8 & 8408.7 & 390.2 & 10148.1 & 498.9 \\
StripeTest APIKEY & 8117.0 & 317.6 & 9893.9 & 384.7 & 8095.1 & 338.0 & 9429.2 & 433.5 & 7927.4 & 339.7 & 9318.6 & 439.4 \\
\bottomrule
\end{tabular}
}
\end{table*}

\begin{table*}[t]
\centering
\small
\caption{Total inference latency (seconds) under \textbf{var-rename} perturbation. Column pairs correspond to DESEC and our method for each victim-to-proxy model transfer.}
\label{tab:latency_var_rename}
\resizebox{\linewidth}{!}{
\begin{tabular}{lcccccccccccc}
\toprule
APIKEY Type & \multicolumn{2}{c}{Deepseek$\rightarrow$Starcode} & \multicolumn{2}{c}{Deepseek$\rightarrow$Seedcode} & \multicolumn{2}{c}{Llama3$\rightarrow$Starcode} & \multicolumn{2}{c}{Llama3$\rightarrow$Seedcode} & \multicolumn{2}{c}{Qwen$\rightarrow$Starcode} & \multicolumn{2}{c}{Qwen$\rightarrow$Seedcode}\\
\cmidrule(lr){2-3}\cmidrule(lr){4-5}\cmidrule(lr){6-7}\cmidrule(lr){8-9}\cmidrule(lr){10-11}\cmidrule(lr){12-13}
& DESEC & Ours & DESEC & Ours & DESEC & Ours & DESEC & Ours & DESEC & Ours & DESEC & Ours \\
\midrule
Deepseek APIKEY & 7483.1 & 219.7 & 7602.7 & 270.0 & 7217.5 & 212.2 & 8213.5 & 260.9 & 7626.8 & 219.1 & 8377.0 & 312.1 \\
Google APIKEY & 6825.2 & 177.0 & 7703.5 & 230.1 & 7129.7 & 167.1 & 7773.7 & 237.7 & 7078.0 & 194.5 & 7613.9 & 219.2 \\
OpenAI APIKEY & 10058.5 & 422.4 & 12190.3 & 514.6 & 9623.3 & 378.3 & 11805.7 & 506.7 & 9999.9 & 399.6 & 12627.0 & 531.3 \\
OpenRouter APIKEY & 8323.0 & 438.2 & 10251.7 & 473.0 & 8530.6 & 430.4 & 10147.3 & 472.7 & 8128.8 & 421.7 & 10536.7 & 491.4 \\
StripeTest APIKEY & 8016.6 & 376.5 & 9354.7 & 408.7 & 7726.4 & 336.5 & 9529.8 & 414.3 & 8366.2 & 337.4 & 9702.5 & 432.2 \\
\bottomrule
\end{tabular}
}
\end{table*}
\clearpage

\subsection{Detailed Results on Secret Recovery across Perturbation Types}
\label{sec:detailed-recovery}
Tables~\ref{tab:rs_commit_edit}--\ref{tab:rs_var_rename} present detailed secret extraction results for five code perturbation tasks. We adopt \textit{Real Rate (RR)} = $\text{RS} / \text{PS}$ as the core metric, where $\text{RS}$ and $\text{PS}$ denote the number of recovered real secrets and generated plausible secret candidates, respectively. RR reflects the proportion of valid genuine API keys among all candidate secrets. We comprehensively compare our method with the DESEC baseline on three victim code models. Gray rows denote our approach, and the parenthesized percentages represent the relative performance gains over the DESEC baseline.

\begin{table}[ht]
\centering
\caption{RR on \textbf{commit\_edit} perturbation task.}
\label{tab:rs_commit_edit}
\setlength{\tabcolsep}{4.0pt}
\renewcommand{\arraystretch}{1.08}
\small
\begin{tabular}{llrrrrr}
\toprule
\textbf{Victim model} & \textbf{Method} & \textbf{DeepSeek} & \textbf{Google} & \textbf{OpenAI} & \textbf{OpenRouter} & \textbf{StripeTest} \\
\midrule
\multirow{4}{*}{DeepSeek-Coder-7B}
& DESEC--StarCoder2-7B & 0.0914 & 0.1073 & 0.0072 & 0.0655 & 0.0055 \\
& DESEC--Seed-Coder-8B & 0.0983 & 0.1130 & 0.0091 & 0.0699 & 0.0058 \\
& \cellcolor{lightgray}Ours--StarCoder2-7B & \cellcolor{lightgray}\textbf{0.1408} & \cellcolor{lightgray}\textbf{0.2169} & \cellcolor{lightgray}\textbf{0.0120} & \cellcolor{lightgray}\textbf{0.0790} & \cellcolor{lightgray}\textbf{0.0081} \\
& \cellcolor{lightgray}Ours--Seed-Coder-8B & \cellcolor{lightgray}\textbf{0.1430} & \cellcolor{lightgray}\textbf{0.1996} & \cellcolor{lightgray}\textbf{0.0114} & \cellcolor{lightgray}\textbf{0.0726} & \cellcolor{lightgray}\textbf{0.0082} \\
\midrule
\multirow{4}{*}{Qwen2.5-Coder-7B}
& DESEC--StarCoder2-7B & 0.0746 & 0.0960 & 0.0204 & 0.0582 & 0.0288 \\
& DESEC--Seed-Coder-8B & 0.0764 & 0.1005 & 0.0199 & 0.0604 & 0.0289 \\
& \cellcolor{lightgray}Ours--StarCoder2-7B & \cellcolor{lightgray}\textbf{0.1038} & \cellcolor{lightgray}\textbf{0.1874} & \cellcolor{lightgray}\textbf{0.0242} & \cellcolor{lightgray}\textbf{0.0664} & \cellcolor{lightgray}\textbf{0.0326} \\
& \cellcolor{lightgray}Ours--Seed-Coder-8B & \cellcolor{lightgray}\textbf{0.1039} & \cellcolor{lightgray}\textbf{0.1850} & \cellcolor{lightgray}\textbf{0.0257} & \cellcolor{lightgray}\textbf{0.0713} & \cellcolor{lightgray}\textbf{0.0319} \\
\midrule
\multirow{4}{*}{LLaMA-3-8B}
& DESEC--StarCoder2-7B & 0.0854 & 0.0990 & 0.0246 & 0.0720 & 0.0847 \\
& DESEC--Seed-Coder-8B & 0.0953 & 0.1087 & 0.0262 & 0.0776 & 0.0861 \\
& \cellcolor{lightgray}Ours--StarCoder2-7B & \cellcolor{lightgray}\textbf{0.1338} & \cellcolor{lightgray}\textbf{0.2145} & \cellcolor{lightgray}\textbf{0.0332} & \cellcolor{lightgray}\textbf{0.0921} & \cellcolor{lightgray}\textbf{0.0956} \\
& \cellcolor{lightgray}Ours--Seed-Coder-8B & \cellcolor{lightgray}\textbf{0.1308} & \cellcolor{lightgray}\textbf{0.2137} & \cellcolor{lightgray}\textbf{0.0334} & \cellcolor{lightgray}\textbf{0.0868} & \cellcolor{lightgray}\textbf{0.0981} \\
\bottomrule
\end{tabular}
\end{table}

\begin{table}[ht]
\centering
\caption{RR on \textbf{dead\_code} perturbation task.}
\label{tab:rs_dead_code}
\setlength{\tabcolsep}{4.0pt}
\renewcommand{\arraystretch}{1.08}
\small
\begin{tabular}{llrrrrr}
\toprule
\textbf{Victim model} & \textbf{Method} & \textbf{DeepSeek} & \textbf{Google} & \textbf{OpenAI} & \textbf{OpenRouter} & \textbf{StripeTest} \\
\midrule
\multirow{4}{*}{DeepSeek-Coder-7B}
& DESEC--StarCoder2-7B & 0.0978 & 0.1040 & 0.0080 & 0.0688 & 0.0061 \\
& DESEC--Seed-Coder-8B & 0.0968 & 0.1020 & 0.0096 & 0.0671 & 0.0058 \\
& \cellcolor{lightgray}Ours--StarCoder2-7B & \cellcolor{lightgray}\textbf{0.1872} & \cellcolor{lightgray}\textbf{0.2053} & \cellcolor{lightgray}\textbf{0.0125} & \cellcolor{lightgray}\textbf{0.0753} & \cellcolor{lightgray}\textbf{0.0086} \\
& \cellcolor{lightgray}Ours--Seed-Coder-8B & \cellcolor{lightgray}\textbf{0.1819} & \cellcolor{lightgray}\textbf{0.2119} & \cellcolor{lightgray}\textbf{0.0123} & \cellcolor{lightgray}\textbf{0.0736} & \cellcolor{lightgray}\textbf{0.0082} \\
\midrule
\multirow{4}{*}{Qwen2.5-Coder-7B}
& DESEC--StarCoder2-7B & 0.0712 & 0.0896 & 0.0212 & 0.0590 & 0.0285 \\
& DESEC--Seed-Coder-8B & 0.0749 & 0.0937 & 0.0203 & 0.0613 & 0.0302 \\
& \cellcolor{lightgray}Ours--StarCoder2-7B & \cellcolor{lightgray}\textbf{0.1414} & \cellcolor{lightgray}\textbf{0.1621} & \cellcolor{lightgray}\textbf{0.0262} & \cellcolor{lightgray}\textbf{0.0689} & \cellcolor{lightgray}\textbf{0.0336} \\
& \cellcolor{lightgray}Ours--Seed-Coder-8B & \cellcolor{lightgray}\textbf{0.1315} & \cellcolor{lightgray}\textbf{0.1689} & \cellcolor{lightgray}\textbf{0.0267} & \cellcolor{lightgray}\textbf{0.0707} & \cellcolor{lightgray}\textbf{0.0354} \\
\midrule
\multirow{4}{*}{LLaMA-3-8B}
& DESEC--StarCoder2-7B & 0.0948 & 0.1007 & 0.0268 & 0.0760 & 0.0853 \\
& DESEC--Seed-Coder-8B & 0.0939 & 0.1029 & 0.0270 & 0.0792 & 0.0881 \\
& \cellcolor{lightgray}Ours--StarCoder2-7B & \cellcolor{lightgray}\textbf{0.1804} & \cellcolor{lightgray}\textbf{0.2010} & \cellcolor{lightgray}\textbf{0.0338} & \cellcolor{lightgray}\textbf{0.0862} & \cellcolor{lightgray}\textbf{0.1037} \\
& \cellcolor{lightgray}Ours--Seed-Coder-8B & \cellcolor{lightgray}\textbf{0.1568} & \cellcolor{lightgray}\textbf{0.2225} & \cellcolor{lightgray}\textbf{0.0347} & \cellcolor{lightgray}\textbf{0.0894} & \cellcolor{lightgray}\textbf{0.0961} \\
\bottomrule
\end{tabular}
\end{table}

\begin{table}[ht]
\centering
\caption{RR on \textbf{format} perturbation task.}
\label{tab:rs_format}
\setlength{\tabcolsep}{4.0pt}
\renewcommand{\arraystretch}{1.08}
\small
\begin{tabular}{llrrrrr}
\toprule
\textbf{Victim model} & \textbf{Method} & \textbf{DeepSeek} & \textbf{Google} & \textbf{OpenAI} & \textbf{OpenRouter} & \textbf{StripeTest} \\
\midrule
\multirow{4}{*}{DeepSeek-Coder-7B}
& DESEC--StarCoder2-7B & 0.1181 & 0.1131 & 0.0081 & 0.0742 & 0.0057 \\
& DESEC--Seed-Coder-8B & 0.1125 & 0.1077 & 0.0096 & 0.0685 & 0.0057 \\
& \cellcolor{lightgray}Ours--StarCoder2-7B & \cellcolor{lightgray}\textbf{0.3686} & \cellcolor{lightgray}\textbf{0.2819} & \cellcolor{lightgray}\textbf{0.0125} & \cellcolor{lightgray}\textbf{0.0891} & \cellcolor{lightgray}\textbf{0.0090} \\
& \cellcolor{lightgray}Ours--Seed-Coder-8B & \cellcolor{lightgray}\textbf{0.3342} & \cellcolor{lightgray}\textbf{0.2535} & \cellcolor{lightgray}\textbf{0.0120} & \cellcolor{lightgray}\textbf{0.0826} & \cellcolor{lightgray}\textbf{0.0086} \\
\midrule
\multirow{4}{*}{Qwen2.5-Coder-7B}
& DESEC--StarCoder2-7B & 0.0740 & 0.0961 & 0.0207 & 0.0602 & 0.0286 \\
& DESEC--Seed-Coder-8B & 0.0872 & 0.1051 & 0.0208 & 0.0606 & 0.0285 \\
& \cellcolor{lightgray}Ours--StarCoder2-7B & \cellcolor{lightgray}\textbf{0.2673} & \cellcolor{lightgray}\textbf{0.2139} & \cellcolor{lightgray}\textbf{0.0260} & \cellcolor{lightgray}\textbf{0.0792} & \cellcolor{lightgray}\textbf{0.0362} \\
& \cellcolor{lightgray}Ours--Seed-Coder-8B & \cellcolor{lightgray}\textbf{0.2717} & \cellcolor{lightgray}\textbf{0.2417} & \cellcolor{lightgray}\textbf{0.0257} & \cellcolor{lightgray}\textbf{0.0787} & \cellcolor{lightgray}\textbf{0.0352} \\
\midrule
\multirow{4}{*}{LLaMA-3-8B}
& DESEC--StarCoder2-7B & 0.1035 & 0.1108 & 0.0273 & 0.0756 & 0.0867 \\
& DESEC--Seed-Coder-8B & 0.1014 & 0.1168 & 0.0283 & 0.0790 & 0.0867 \\
& \cellcolor{lightgray}Ours--StarCoder2-7B & \cellcolor{lightgray}\textbf{0.2994} & \cellcolor{lightgray}\textbf{0.2545} & \cellcolor{lightgray}\textbf{0.0334} & \cellcolor{lightgray}\textbf{0.0938} & \cellcolor{lightgray}\textbf{0.1081} \\
& \cellcolor{lightgray}Ours--Seed-Coder-8B & \cellcolor{lightgray}\textbf{0.3098} & \cellcolor{lightgray}\textbf{0.2620} & \cellcolor{lightgray}\textbf{0.0354} & \cellcolor{lightgray}\textbf{0.0916} & \cellcolor{lightgray}\textbf{0.1001} \\
\bottomrule
\end{tabular}
\end{table}

\begin{table}[ht]
\centering
\caption{RR on \textbf{syntax\_equiv} perturbation task.}
\label{tab:rs_syntax_equiv}
\setlength{\tabcolsep}{4.0pt}
\renewcommand{\arraystretch}{1.08}
\small
\begin{tabular}{llrrrrr}
\toprule
\textbf{Victim model} & \textbf{Method} & \textbf{DeepSeek} & \textbf{Google} & \textbf{OpenAI} & \textbf{OpenRouter} & \textbf{StripeTest} \\
\midrule
\multirow{4}{*}{DeepSeek-Coder-7B}
& DESEC--StarCoder2-7B & 0.0945 & 0.0994 & 0.0076 & 0.0675 & 0.0058 \\
& DESEC--Seed-Coder-8B & 0.0959 & 0.1003 & 0.0091 & 0.0692 & 0.0056 \\
& \cellcolor{lightgray}Ours--StarCoder2-7B & \cellcolor{lightgray}\textbf{0.3247} & \cellcolor{lightgray}\textbf{0.3032} & \cellcolor{lightgray}\textbf{0.0129} & \cellcolor{lightgray}\textbf{0.0852} & \cellcolor{lightgray}\textbf{0.0091} \\
& \cellcolor{lightgray}Ours--Seed-Coder-8B & \cellcolor{lightgray}\textbf{0.2849} & \cellcolor{lightgray}\textbf{0.2647} & \cellcolor{lightgray}\textbf{0.0123} & \cellcolor{lightgray}\textbf{0.0778} & \cellcolor{lightgray}\textbf{0.0086} \\
\midrule
\multirow{4}{*}{Qwen2.5-Coder-7B}
& DESEC--StarCoder2-7B & 0.0691 & 0.0897 & 0.0198 & 0.0567 & 0.0284 \\
& DESEC--Seed-Coder-8B & 0.0716 & 0.0898 & 0.0189 & 0.0562 & 0.0287 \\
& \cellcolor{lightgray}Ours--StarCoder2-7B & \cellcolor{lightgray}\textbf{0.2215} & \cellcolor{lightgray}\textbf{0.2353} & \cellcolor{lightgray}\textbf{0.0257} & \cellcolor{lightgray}\textbf{0.0683} & \cellcolor{lightgray}\textbf{0.0337} \\
& \cellcolor{lightgray}Ours--Seed-Coder-8B & \cellcolor{lightgray}\textbf{0.2305} & \cellcolor{lightgray}\textbf{0.2370} & \cellcolor{lightgray}\textbf{0.0265} & \cellcolor{lightgray}\textbf{0.0724} & \cellcolor{lightgray}\textbf{0.0345} \\
\midrule
\multirow{4}{*}{LLaMA-3-8B}
& DESEC--StarCoder2-7B & 0.0838 & 0.0961 & 0.0249 & 0.0737 & 0.0834 \\
& DESEC--Seed-Coder-8B & 0.0840 & 0.1011 & 0.0252 & 0.0775 & 0.0827 \\
& \cellcolor{lightgray}Ours--StarCoder2-7B & \cellcolor{lightgray}\textbf{0.2780} & \cellcolor{lightgray}\textbf{0.2768} & \cellcolor{lightgray}\textbf{0.0337} & \cellcolor{lightgray}\textbf{0.0897} & \cellcolor{lightgray}\textbf{0.1061} \\
& \cellcolor{lightgray}Ours--Seed-Coder-8B & \cellcolor{lightgray}\textbf{0.2577} & \cellcolor{lightgray}\textbf{0.2627} & \cellcolor{lightgray}\textbf{0.0353} & \cellcolor{lightgray}\textbf{0.0944} & \cellcolor{lightgray}\textbf{0.0993} \\
\bottomrule
\end{tabular}
\end{table}

\begin{table}[ht]
\centering
\caption{RR on \textbf{var\_rename} perturbation task.}
\label{tab:rs_var_rename}
\setlength{\tabcolsep}{4.0pt}
\renewcommand{\arraystretch}{1.08}
\small
\begin{tabular}{llrrrrr}
\toprule
\textbf{Victim model} & \textbf{Method} & \textbf{DeepSeek} & \textbf{Google} & \textbf{OpenAI} & \textbf{OpenRouter} & \textbf{StripeTest} \\
\midrule
\multirow{4}{*}{DeepSeek-Coder-7B}
& DESEC--StarCoder2-7B & 0.0883 & 0.0936 & 0.0073 & 0.0624 & 0.0058 \\
& DESEC--Seed-Coder-8B & 0.0846 & 0.0950 & 0.0085 & 0.0641 & 0.0057 \\
& \cellcolor{lightgray}Ours--StarCoder2-7B & \cellcolor{lightgray}\textbf{0.1541} & \cellcolor{lightgray}\textbf{0.1518} & \cellcolor{lightgray}\textbf{0.0112} & \cellcolor{lightgray}\textbf{0.0825} & \cellcolor{lightgray}\textbf{0.0086} \\
& \cellcolor{lightgray}Ours--Seed-Coder-8B & \cellcolor{lightgray}\textbf{0.1432} & \cellcolor{lightgray}\textbf{0.1429} & \cellcolor{lightgray}\textbf{0.0111} & \cellcolor{lightgray}\textbf{0.0738} & \cellcolor{lightgray}\textbf{0.0083} \\
\midrule
\multirow{4}{*}{Qwen2.5-Coder-7B}
& DESEC--StarCoder2-7B & 0.0631 & 0.0874 & 0.0173 & 0.0538 & 0.0274 \\
& DESEC--Seed-Coder-8B & 0.0646 & 0.0881 & 0.0186 & 0.0560 & 0.0276 \\
& \cellcolor{lightgray}Ours--StarCoder2-7B & \cellcolor{lightgray}\textbf{0.1156} & \cellcolor{lightgray}\textbf{0.1342} & \cellcolor{lightgray}\textbf{0.0231} & \cellcolor{lightgray}\textbf{0.0679} & \cellcolor{lightgray}\textbf{0.0340} \\
& \cellcolor{lightgray}Ours--Seed-Coder-8B & \cellcolor{lightgray}\textbf{0.1144} & \cellcolor{lightgray}\textbf{0.1344} & \cellcolor{lightgray}\textbf{0.0242} & \cellcolor{lightgray}\textbf{0.0718} & \cellcolor{lightgray}\textbf{0.0338} \\
\midrule
\multirow{4}{*}{LLaMA-3-8B}
& DESEC--StarCoder2-7B & 0.0809 & 0.0953 & 0.0239 & 0.0718 & 0.0855 \\
& DESEC--Seed-Coder-8B & 0.0790 & 0.0988 & 0.0236 & 0.0727 & 0.0814 \\
& \cellcolor{lightgray}Ours--StarCoder2-7B & \cellcolor{lightgray}\textbf{0.1370} & \cellcolor{lightgray}\textbf{0.1539} & \cellcolor{lightgray}\textbf{0.0295} & \cellcolor{lightgray}\textbf{0.0896} & \cellcolor{lightgray}\textbf{0.0985} \\
& \cellcolor{lightgray}Ours--Seed-Coder-8B & \cellcolor{lightgray}\textbf{0.1284} & \cellcolor{lightgray}\textbf{0.1522} & \cellcolor{lightgray}\textbf{0.0306} & \cellcolor{lightgray}\textbf{0.0889} & \cellcolor{lightgray}\textbf{0.0958} \\
\bottomrule
\end{tabular}
\end{table}

\end{document}